\documentclass[12pt]{article}
\usepackage{amsfonts,amsmath,amssymb,epsf,cite}
\usepackage{graphicx,color}
\usepackage[usenames,dvipsnames]{pstricks}
 \def\ep{{\epsilon}}

 \def\frac#1#2{{#1\over #2}}

 \def\s{\sqrt}

 \def\de{\partial}

 \def\f {\frac}
 \def\ti{\tilde}
 \def\ap{\alpha}

 \def\ddd{\cdot\cdot\cdot}
 \def\no{\nonumber \\}

 \def\la{\langle}
 \def\lb{\rangle}
 \def\ep{\epsilon}

 \def\ep{{\epsilon}}

 \def\frac#1#2{{#1\over #2}}

 \def\s{\sqrt}

\def\be{\begin{equation}}
\def\ee{\end{equation}}
\def\ba{\begin{eqnarray}}
\def\ea{\end{eqnarray}}

\begin{document}
\begin{titlepage}
\thispagestyle{empty}
\begin{flushright}
IPMU11-0180
\end{flushright}

\bigskip

\begin{center}
\noindent{\large \textbf{Holographic Fermi Surfaces and Entanglement Entropy}}\\
\vspace{2cm}

\noindent{Noriaki
Ogawa\footnote{e-mail: noriaki.ogawa@ipmu.jp}, Tadashi
Takayanagi\footnote{e-mail:tadashi.takayanagi@ipmu.jp} and Tomonori
Ugajin\footnote{e-mail:tomonori.ugajin@ipmu.jp}}

\vspace{1cm}
  {\it
Institute for the Physics and Mathematics of the Universe (IPMU), \\
 University of Tokyo, Kashiwa, Chiba 277-8582, Japan\\
 }
\end{center}
\vspace{1cm}
\begin{abstract}
We argue that Landau-Fermi liquids do not have any gravity duals
in the purely classical limit. We employ the logarithmic
behavior of entanglement entropy to characterize the existence of Fermi surfaces.
By imposing the null energy condition, we show that the specific heat always
behaves anomalously. We also present a classical gravity dual which
has the expected behavior of the entanglement entropy and specific
heat for non-Fermi liquids.
\end{abstract}
\end{titlepage}
\newpage

\section{Introduction and Summary}

The AdS/CFT correspondence \cite{Maldacena} provides us a very
useful tool to study various properties of strongly coupled systems.
It is clear that among many possible states in condensed matter
physics, systems with Fermi surfaces are remarkably important.
Recently, there have been exciting progresses on the holographic
realizations of Fermi surfaces using the AdS/CFT such as the ones
using the Reissner-Nordstr\"om black holes \cite{MIT,Za,Rey,MITR} and
the star-like solutions, called electron stars \cite{ESa,ESb,ESc}.
 Refer also to \cite{SaF,ILM,GSW,SaH} for further developments and
 other constructions. The former
construction leads to the non-Fermi liquids, while the latter to the
standard Landau-Fermi liquids. The non-Fermi liquids have anomalous
behaviors, for example, with respect to the specific heat and
conductivity when we compare them with the Landau-Fermi liquids.

An outstanding advantage of AdS/CFT is that we can analyze strongly
coupled systems in a classical gravity by ignoring all quantum
corrections. However, there have not been found any holographic
constructions of Fermi surfaces where basic quantities such as the
free energy, the entropy and the specific heat are all calculated in
the classical gravity and agree with the expected standard behaviors
of Fermi surfaces. For example, a charged black hole whose near
horizon includes AdS$_2$ has the unusual property of the
non-vanishing entropy even at zero temperature. On the other hand,
in the electron star solutions, which approach to the Lifshitz
geometry \cite{KLM} in the IR limit, the specific heat and entropy
are not large enough to be calculated in the classical gravity. The
purpose of this paper is to study if it is possible to construct a
holographic dual of Fermi surfaces in a purely classical gravity. If
such a gravity dual exists, we expect that the number of Fermi surfaces 
is of order $N^2$, taking the Fermi energy to be order one. Here we are considering
a system where fermions belong to the adjoint representation of the $SU(N)$ gauge group. 
The fractionalized fermi liquid \cite{SaF} is a promising candidate of
such $O(N^2)$ fermi surfaces.

For this purpose, we need to make the definition of Fermi surfaces
clear. In this paper, we employ the entanglement entropy
\cite{BoSr,EIR,CaCaR,CaHuR} to characterize the presence of Fermi
surface. We define systems with Fermi surfaces by requiring
that {\it their entanglement entropies show the logarithmic violation of
the area law}. This has been shown explicitly for free fermion
theories \cite{EEF} and recent progresses in condensed matter physics
support that this property is true even
for non-Fermi liquids \cite{SW,spin}.

By using the holographic entanglement entropy \cite{RT,RTL,RTR}, we
can rewrite this requirement as the IR behavior of the metric in the
gravity dual. We mainly focus on 2+1 dimensional systems dual to
asymptotic AdS$_4$ geometries. Then we require the null energy condition, which
is expected to be satisfied by any physically sensible backgrounds with
well-defined holographic duals. This leads to additional constraints on
the metric. Eventually, we find that to satisfy these two
requirements the specific heat $C$ should behave like $C\propto
T^\ap$ with the constraint $\ap\leq 2/3$ at low temperature $T$ for
2+1 dimensional systems.
Thus the standard Fermi liquid ($\ap=1$) is not
allowed. In this way, we find that under our classical gravity dual
assumption, we can construct only non-Fermi liquids. We also give an
example of an effective gravity action which has such a gravity
solution. We will leave the embeddings of our gravity backgrounds in
string theory as a future problem.

This paper is organized as follows: In section two, we discuss the
behavior of entanglement entropy in the presence of Fermi surfaces.
In section three, we study the constraints on the gravity dual
metric for the Fermi surfaces from the viewpoint of the entanglement
entropy. In section four, we further impose the null energy
condition and see how it constrains the behavior of the specific
heat. In section five, we present an effective gravity model which
has the expected behavior of the entanglement entropy and specific
heat for non-Fermi liquids.

\section{Fermi Surface and Entanglement Entropy}

Consider the entanglement entropy in a $d+1$ dimensional quantum field theory (QFT) on $R^{1,d}$.
The coordinate of $R^{1,d}$ is denoted by $(t,x_1,x_2,\ddd, x_d)$. The entanglement entropy
$S_A$ associated with a subsystem $A$ is defined by the von-Neumann entropy
$S_A=-\mbox{Tr}\rho_A\log \rho_A$. $\rho_A$ is the reduced density matrix defined in terms of the
total density matrix $\rho$ by $\rho_A=\mbox{Tr}_B\rho$, where the subsystem $B$ is the complement of $A$.

We choose the subsystem $A$ as a strip with the width $l$. This is explicitly
expressed as
\be
A=\{(x_1,x_2,\ddd,x_d)|-\f{l}{2}\leq x_1\leq \f{l}{2},\ \ \ 0\leq x_2,x_3,\ddd,x_d\leq L\}.
\label{suba}
\ee

In a $d+1$ dimensional conformal field theory, the entanglement
entropy $S_A$ behaves like \be S_A=\gamma\f{L^{d-1}}{\ep^{d-1}}-\ap
\f{L^{d-1}}{l^{d-1}}, \ee where $\ep$ is the UV cut off ($\ep\to 0$
in the UV limit); $\gamma$ and $\ap$ are numerical constants which
represent the degrees of freedom \cite{CaCaR,CaHuR}. This has been
confirmed in the holographic calculation \cite{RT,RTL}. The leading
divergent term is proportional to the area of the boundary of $A$
and this is called the area law \cite{BoSr,EIR}. This is true for
all QFTs with UV fixed points.

\subsection{Logarithmic Violation of Area Law}

On the other hand, for a $d+1$ dimensional QFT with a Fermi surface,
we expect that the finite part of $S_A$ is substantially modified
when the size $l$ of the subsystem $A$ is large enough \be
S_A=\gamma\f{L^{d-1}}{\ep^{d-1}}+\eta L^{d-1}k_F^{d-1}\log
(lk_F)+O(l^0),\label{fermiee} \ee where $k_F$ is the Fermi momentum
and $\eta$ is a positive numerical constant.

Before we give a field theoretic explanation of the behavior
(\ref{fermiee}) below, we would like to point out that
(\ref{fermiee}) is essentially the same as the logarithmic violation
of area law in femionic theories, which is well known in the
condensed matter literature.

In the papers \cite{EEF} (see also the review \cite{EIR}), the
violation of area law in free fermions on lattices has been shown in
any dimension $d$. They argue that the leading term in the entropy
looks like \be S_A\sim L^{d-1}\log L+\ddd, \label{vio} \ee where 
we assume that the subsystem $A$ is given by a finite size (but large) 
region (e.g. a d dimensional ball) which spans order $L$ in any directions. 
Their calculation is based on a discretized lattice setup,
and there they assume that the Fermi surface has a finite area. In our continuum limit of the quantum field theories, this
corresponds to the case where the Fermi energy is the same order of
the UV cut off scale i.e. $k_F\sim \ep^{-1}$. Indeed, in this limit,
the second term in (\ref{fermiee}) gets dominant and leads to the
violation of the area law (\ref{vio}) by setting 
$l\sim L$. However, in this paper what
we are interested in is the case where $k_F$ is finite while $\ep$
is taken to be zero. Therefore we have the expression
(\ref{fermiee}) instead of (\ref{vio}).

Even in the presence of strong interactions this kind of the
violation of area law has been expected. The recent calculation of
$S_A$ for a critical spin liquid state \cite{spin} can be regarded
as an evidence of such a log term even in a strongly coupled fermion
theory with a Fermi surface. Also such a violation of area law has
also been argued in \cite{SW}.

\subsection{Free Fermion Calculation}

To understand (\ref{fermiee}), we would like to consider a
calculation of $S_A$  for a free fermion in $(d+1)$
dimension\footnote{Similar estimations have already been done in
\cite{RTR,SW}. See also \cite{DSY} for an explicit calculation.}.
We compactify $x_1,x_2,\ddd,x_d$ on a torus with a
large radius $L$. The momentum in these directions are quantized as
\be k_i=\f{n^i}{L},\ \ \ \ (i=1,2,\ddd,d). \ee We assume there
exists a Fermi surface with the Fermi momentum $k_F$ and the ground
state is describes as \be
|\Psi\lb=\prod_{|k|<k_F}b^{\dagger}_{n_1,n_2,\ddd,n_d}|0\lb. \ee

Since the theory is free, we can treat sectors with different
quantum numbers $\vec{n}\equiv(n_2,n_3,\ddd,n_d)$ as decoupled
independent sectors. Each sector with a fixed number of $\vec{n}$
can be regarded as a $1+1$ dimensional free massive fermion theory
with the Fermi surface at $k_1=\s{k_F^2-\vec{k}^2}$. The mass is
given by $m=|\vec{k}|$. The ground state in the $1+1$ dimensional
theory for a fixed $\vec{n}$ looks like \be
|\Psi_{\vec{n}}\lb=\prod_{k_1\leq
\s{k_F^2-|\vec{k}|^2}}b^{\dagger}_{n_1,\vec{n}}|0\lb, \ee and the
total ground state is given by \be
|\Psi\lb=\prod_{\vec{n}}|\Psi_{\vec{n}}\lb. \ee Notice that when
$k_F<|\vec{k}|$ there is no Fermi surface in the reduced two
dimensional fermion theory.

Since the density matrix is factorized into sectors with various
$\vec{n}$, we can express $S_A$ as the summation over each sector:
\be S^{(d+1)}_A=\sum_{\vec{n}}S^{(1+1)}_A(\vec{n}). \ee

We are interested in the region \be l>>\f{1}{k_F}, \ee so that the
entanglement entropy can probe the IR physics. When $|\vec{k}|>k_F$,
there is no Fermi surface and $S^{(1+1)}_A(\vec{n})$ can be found
from the known result in two dimensional CFT. If we define
$S^{QFT_2}_A$ to be the entanglement entropy for a massive two
dimensional QFT without any Fermi surface, then it behaves like \ba
&& S^{QFT_2}_A\sim \log l/\ep \ \ (lm<<1),\no && S^{QFT_2}_A\sim
-\log m\ep \ \ \ (lm>>1).\label{becft} \ea Note also that we omitted
the coefficient, which is proportional to the central charge
\cite{CaCaR}.

Since $l>>\f{1}{k_F}$, when $|\vec{k}|>k_F$ we find \be
S^{(1+1)}_A(\vec{n})\sim -\log (|\vec{k}|\ep). \ee

In this way we we obtain the following estimation \ba && S^{(d+1)}_A
\sim L^{d-1}\int_{k_F<|\vec{k}|<\ep^{-1}}
(dk)^{d-1}S^{F}_A\left(|\vec{k}|,0\right)+L^{d-1}\int_{|\vec{k}|<k_F}
(dk)^{d-1}S^{F}_A\left(|\vec{k}|,\s{k_F^2-|\vec{k}|^2}\right),\no &&
\sim  -L^{d-1}\int_{k_F<|\vec{k}|<\ep^{-1}} (dk)^{d-1} \log
(\ep|\vec{k}|) +L^{d-1}\int_{|\vec{k}|<k_F}
(dk)^{d-1}S^{F}_A\left(|\vec{k}|,\s{k_F^2-|\vec{k}|^2}\right),
\label{intee} \ea where $S^F_{A}(m,\mu)$ is the entanglement entropy
for a free massive fermion in $1+1$ dimension with the mass $m$ and
the Fermi momentum $\mu$. The first term in (\ref{intee}) comes from
the modes $|\vec{k}|>k_F$ can be simplify estimated as \be
-L^{d-1}\int_{k_F<|\vec{k}|<\ep^{-1}} (dk)^{d-1}\log
(\ep|\vec{k}|)\sim \left(\f{L}{\ep}\right)^{d-1}+\ddd, \label{intar} \ee which just
represents the standard area law.

To proceed further we need to know $S_A^F$ for non-zero values of
$\mu$. The dispersion relation in the massive two dimensional
fermion obtained by the dimensional reduction looks like \ba
E_k&=&\s{|\vec{k}|^2+\left(\s{k_F^2-|\vec{k}|^2}+\delta
k_1\right)^2},\no &\simeq & k_F+\f{\s{k_F^2-|\vec{k}|^2}}{k_F}\delta
k_1. \label{rela} \ea Therefore near the Fermi surface (i.e. $\delta
k_1<<k_F$), the fermion behaves like a massless fermion. Therefore
we can estimate the remained part of $S_A$ as follows \ba &&
L^{d-1}\int_{|\vec{k}|<k_F}
(dk)^{d-1}S^{F}_A\left(|\vec{k}|,\s{k_F^2-|\vec{k}|^2}\right)\no &&
\sim L^{d-1}\int_{|\vec{k}|<k_F} (dk)^{d-1} \log l/\ep.  \label{eefe} \ea
This is because the
low energy excitation around the Fermi surface becomes relativistic as in
(\ref{rela}).

Therefore, by combining (\ref{intar}) and (\ref{eefe}),
 we find that the total behavior of $S_A$
looks like \be S^{(d+1)}_A\sim
\left(\f{L}{\ep}\right)^{d-1}+L^{d-1}k_F^{d-1}\log (lk_F)+\ddd. \ee
Notice that the logarithmic divergent terms $O(\log \ep)$ are obviously
canceled between (\ref{intar}) and (\ref{eefe}). In this way, we managed to
derive the formula (\ref{fermiee}).

So far we assumed the free fermions. In order to compare the results
with those in AdS/CFT, we need to see how this argument changes due
to the strong interactions. However, there are evidences which argue
that the logarithmic dependence (\ref{fermiee}) does not change. One
of them is the recent calculations done for the interacting Fermi
surfaces in spin liquids \cite{spin}. Another is the fact that
though in non-Fermi liquids we often encounter non-relativistic
dynamical exponents for the dispersion relation around the Fermi
surface, the logarithmic behavior (\ref{becft}) does not change
(only its coefficient changes). Therefore, in this paper, we define
systems with Fermi surfaces by requiring that their entanglement
entropies show the logarithmic violation of the area law, as already
mentioned.

\subsection{Entanglement Entropy for 2D Free Massless Dirac Fermions}

It is useful to look at an analytical expression of the entanglement
entropy in two dimensional CFT at finite chemical potential. Below
we will calculate $S_A$ for a two dimensional free massless Dirac
fermion on a circle at finite temperature $T=1/\beta$ and at finite
chemical potential $\mu$. We would like to ask the readers to refer
to \cite{ANT} for details where a similar analysis has been done
when $\mu=0$.

We consider a torus whose lengths in the space and the Euclidean
time direction are given by $\beta$ and $1$, respectively. The size
of subsystem $A$ is denoted by $L$ such that $0<L<1$. The thermal
partition function is expressed by using the theta functions \be
Z_{th}=\mbox{Tr}e^{-\beta H-2\pi\beta \mu N}
=\f{|\theta_3(i\beta\mu|i\beta)|^2}{|\eta(i\beta)|^2}, \ee
 where $H$ and $N$ are Hamiltonian and the fermion number.

The two point function of twisted operators $\sigma_k$ in the $Z_N$
orbifolded free fermion theory on a torus is given by \be \la
\sigma_k(L)\sigma_{-k}(0)\lb=
\left|\f{2\pi\eta(i\beta)^3}{\theta_1(L|i\beta)}
\right|^{4\Delta_k}\cdot
\left|\f{\theta_3(\f{kL}{N}+i\beta\mu|i\beta)}{\theta_3(i\beta\mu|i\beta)}\right|^2,
\ee where $\Delta_k=\f{k^2}{2N^2}$ is the conformal dimension of the
$k$-th twisted operator $\sigma_{k}$. By applying the replica trick,
we obtain \be S_A=-\f{\de}{\de N}\left[\log
\mbox{Tr}[(\rho_A)^n]\right]\Bigr|_{N=1} =-\f{\de}{\de N}\left[\log
\prod^{\f{N-1}{2}}_{k=-\f{N-1}{2}}\la
\sigma_k(L)\sigma_{-k}(0)\lb\right]\Biggr|_{N=1}. \ee In the end, we
find the high temperature expansion \ba
S_A(L,\beta,\mu)&=&\frac{1}{3}\log \left[\frac{\beta}{\pi \ep} \sinh
\left(\frac{\pi L}{\beta} \right)\right]+\frac{1}{3}
\sum^{\infty}_{m=1}\log\left[\frac{(1-e^{2\pi \frac{L}{\beta}}e^{-2\pi\frac{m}{\beta}})(1-e^{-2\pi \frac{L}{\beta}}e^{-2\pi\frac{m}{\beta}})}{(1-e^{-2\pi\frac{m}{\beta}})^2} \right] \nonumber  \\
& + & 2\sum^{\infty}_{l=1}\frac{(-1)^l}{l}\left(\frac{\frac{\pi lL}{\beta}\coth\left( \frac{\pi l L}{\beta}\right) -1}{\sinh \left( \frac{\pi l}{\beta}\right) }\right)\cos (2\pi \mu l ),
\ea
where $\ep(\to 0)$ is the UV cut off.

We find basic properties as follows \be
S_A(L,\beta,\mu)=S_A(L,\beta,1-\mu)=S_A(L,\beta,\mu+1), \ee which is
explained by the particle-hole exchange symmetry and the periodicity
of energy levels peculiar to the two dimensional free fermion
theory.

We can also confirm that the difference \be
S_A(L=1-\delta,\beta,\mu)-S_A(L=\delta,\beta,\mu)
=\frac{\pi}{3\beta}+2\sum^{\infty}_{l=1}\frac{(-1)^l}{l}
\left(\frac{\frac{\pi lL}{\beta}\coth\left( \frac{\pi
l}{\beta}\right) -1} {\sinh \left( \frac{\pi l}{\beta}\right)
}\right)\cos (2\pi \mu l ), \ee in the limit $\delta\to 0$ actually
coincides with the thermal entropy \be
S_{th}(\beta,\mu)=-\beta^2\f{\de}{\de\beta}\left(\beta^{-1}\log
Z_{th}\right). \ee

In the low temperature expansion entanglement entropy is given by
(for $0\leq \mu<1$)
\begin{eqnarray}
S_A(L,\beta,\mu)&=&\frac{1}{3}\log\left[\frac{1}{\pi \ep}\sin(\pi
L)\right] +\frac{1}{3}\sum_{m=1}^{\infty}\log \left[
\frac{(1+e^{2\pi i L}e^{-2\pi \beta (m-1/2)})^2(1+e^{-2\pi i L}
e^{-2\pi \beta (m-1/2)})^2}{(1+e^{-2\pi \beta (m-1/2)})^2}\right] \nonumber \\
& + &2\sum^{\infty}_{l=1}
\frac{(-1)^l}{l}\left(\frac{\pi lL\cot (\pi L l) -1}
{\sinh \left( \pi \beta l\right)}\right)\cosh (2\pi \mu \beta l )
\end{eqnarray}

Thus at zero temperature, we simply find \be
S_A(L,\infty,\mu)=\frac{1}{3}\log\left[\frac{1}{\pi \ep}\sin(\pi
L)\right], \ee we confirm that the logarithmic behavior still exists
even if we turn on the chemical potential as we assumed in the
previous calculations.

\section{Possible Holographic Duals with Fermi Surfaces}

We are interested in holographic duals of strongly coupled $2+1$
dimensional systems with Fermi surfaces. Thus we consider $3+1$
dimensional gravity backgrounds. We consider the following metric \be
ds^2=\f{R^2}{z^2}\left(-f(z)dt^2+g(z)dz^2+dx^2+dy^2\right).
\label{metric} \ee 
This is the most general metric when we require the translational and rotational symmetry in $x$ and $y$ directions.\footnote{Here we did not assume any breaking of symmetry as we are interested
in realizing a fermi liquid in the absence of any external fields. However, it is also
possible that the logarithmic behavior of entanglement entropy occurs in
the presence of external fields. A good example will be the magnetic
quantum critical point found in an asymptotically AdS$_5$ geometry \cite{DhKr}, as pointed out in \cite{SWH}}  Notice that the metric component $g_{tz}$ can be eliminated by a coordinate 
transformation. We require that its dual theory has a UV fixed point
and therefore (\ref{metric}) should be asymptotically AdS$_4$. Since
$z=0$ corresponds to the boundary, we require \be f(0)=g(0)=1.
\label{adscon} \ee

\subsection{Holographic Entanglement Entropy}

Now we holographically calculate the entanglement entropy in these
gravity backgrounds. The holographic entanglement entropy
\cite{RT,RTL,RTR} is given by \be
S_A=\f{\mbox{Area}(\gamma_A)}{4G_N}, \ee where $\gamma_A$ is the
codimension two minimal area surface which coincides with $\de A$ at
the boundary $z=0$.

We are interested in the strip shape subsystem $A$ (\ref{suba}) i.e.
\be A=\{(x,y)|-\f{l}{2}\leq x\leq \f{l}{2},\ \ \ 0\leq y \leq L\}.
\ee We can specify the minimal area surface $\gamma_A$ by
 the surface $x=x(z)$ in (\ref{metric}).

The area of this surface can be found as \be
\mbox{Area}(\gamma_A)=2R^2L\int^{z_*}_\ep\f{dz}{z^2}\s{g(z)+x'(z)^2},
\label{area} \ee where $z_*$ is the turning point, where $x'$ gets
divergent; $\ep$ is the UV cut off. The variational principle for
$x(z)$ leads to \be \f{x'(z)}{z^2\s{g(z)+x'^2}}=\mbox{const.}\ \ \ \
, \ee which can be solved as \be
x'(z)=\f{z^2}{z_*^2}\s{\f{g(z)}{1-\f{z^4}{z_*^4}}}\ \ \ . \ee

The width $l$ is related to $z_*$ by \be
l=2\int^{z_*}_{0}dz\f{z^2}{z_*^2}\s{\f{g(z)}{1-\f{z^4}{z_*^4}}}.
\label{length} \ee In the end, the area (\ref{area}) is expressed as
follows \be
\mbox{Area}(\gamma_A)=2R^2L\int^{z_*}_\ep\f{dz}{z^2}\s{\f{g(z)}{1-\f{z^4}{z_*^4}}}.
\label{areaz} \ee

To obtain explicit results, let us assume that the function $g(z)$
scales as \ba g(z)&\simeq& \left(\f{z}{z_F}\right)^{2n}\ \ \
(z>>z_F), \no &\simeq & \  \ \   1 \ \ \ \ \  \ \ \ (z<<z_F),
\label{bha} \ea with a certain scale $z_F$. The parameter $z_F$ is
dual to the length scale where the geometry starts modified and
shows peculiar IR behaviors. We assume $n>1$ below. We are
interested in the very IR limit $z_*>>z_F$ and would like to see the
$l$ dependence of the finite part of $S_A$.

Then (\ref{length}) is estimated as \be l\sim
\f{2}{z_*^2z_F^n}\int^{z_*}dz \f{z^{2+n}}{\s{1-z^4/z_*^4}}\sim
c_n\f{z_*^{n+1}}{z_F^n}, \label{lint} \ee where $c_n$ is a positive numerical
constant.

On the other hand, the area is evaluated as \ba
\mbox{Area}(\gamma_A)&\simeq &
\f{2R^2L}{\ep}+\f{2R^2L}{z_F^n}\int^{z_*}dz\f{z^{n-2}}{\s{1-z^4/z_*^4}}\no
&\simeq &\f{2R^2L}{\ep}+d_n \f{R^2Lz_*^{n-1}}{z_F^n},
\label{areaint} \ea where $d_n$ is a positive numerical
constant.

In the end, the behavior of $S_A$ is obtained as follows ($k_n$ is a
positive constant) \be
S_A=\f{R^2L}{2G^{(4)}_N\ep}+k_n\f{R^2}{G_N}\f{L}{z_F}\cdot
\left(\f{l}{z_F}\right)^{\f{n-1}{n+1}}+\ddd, \ee where the omitted
term represents the subleading term in the limit $z_*>>z_F$ or
equally $l>>z_F$. The leading divergent term agrees with the area
law and this is expected because our background is asymptotically
AdS. Note that the maximum increasing rate of $S_A$ as a function of
$l$ is linear, which corresponds to $n\to\infty$ limit. This is
consistent with the fact that the maximal allowed entropy is always
given by the log of the dimensional of Hilbert space of $A$, which
is clearly proportional to $l$. This limit is realized in the charge
AdS black holes whose the near horizon geometry AdS$_2\times$R$^2$.
In this case, the spacetime has a horizon with non-vanishing entropy
and the extensive behavior of $S_A$ is indeed expected.

On the other hand, for the purpose of a gravity dual of a Fermi
surfaces, we need to set $n=1$. In this case, we indeed find the
following behavior (notice that (\ref{lint}) remains the same, while
we need to modify (\ref{areaint})) \be
S_A=\f{R^2L}{2G^{(4)}_N\ep}+k_1\f{R^2}{G_N}\f{L}{z_F}\log\left(\f{l}{z_F}\right)+O(l^0),
\ee which agrees with (\ref{fermiee}). Therefore the existence of
Fermi surfaces requires the behavior \ba g(z)&\simeq&
\left(\f{z}{z_F}\right)^{2}\ \ \ (z>>z_F), \no &\simeq & \  \ \   1
\ \ \ \ \  \ \ \ (z<<z_F). \label{fermig} \ea The parameter $z_F$ is
now interpreted as the scale of the Fermi surface $\sim k_F^{-1}$.
More precisely, since we are expecting the existence of
many fermi surfaces, $z_F$ is regarded as the average of their fermi levels.
 This is our basic guiding principle to search a holographic dual of
Fermi surfaces. It is also straightforward to extend the above
argument to higher dimensions.

\subsection{Circular Subsystems}

So far we analytically confirmed the logarithmic dependence on the
subsystem size $l$ only for the strip subsystems. Here we would like
to study the case where the subsystem $A$ is given by a circular
disk with radius $l$. In conformal field theories, we obtain \be
S_A=\gamma\cdot\f{l}{\ep}-\delta, \ee where $\gamma$ and $\delta$
are constants \cite{CaHuR,RT,MFS}. In the presence of the Fermi
surfaces, we again expect \be
S_A=\gamma\cdot\f{l}{\ep}+\ti{\eta}\cdot lk_F \log lk_F+\ddd.
\label{fsll} \ee

In the holographic calculations with the metric (\ref{metric}), we
introduce a polar coordinate such that
$dx^2+dy^2=dr^2+r^2d\theta^2$. The subsystem $A$ of the CFT
corresponds to $r\leq l$ at $z=0$ in the gravity background. The
surface $\gamma_A$ is specified by $r=r(z)$ in (\ref{metric}) at a
fixed time $t$. The holographic entanglement entropy is given by
minimizing \ba S_A=\f{\pi R^2}{2G_N}\int^{z_*}_\ep
\f{dz}{z^2}r(z)\s{g(z)+r'(z)^2}. \ea The boundary condition of
$r(z)$ is $r(0)=l$ and $r(z_*)=0$. At the turning point $z=z_*$,
$r'(z)$ gets divergent. The equation of motion for $r(z)$ is given
by \be
\de_z\left(\f{r(z)r'(z)}{z^2\s{g(z)+r'(z)^2}}\right)=\f{\s{g(z)+r'(z)^2}}{z^2}.
\label{eom} \ee The behavior near $z=z_*$ is completely fixed by the
equation of motion (\ref{eom}) and given in the form \ba r(z)\simeq
(z_*-z)^{1/2}\left(r_0+r_1(z_*-z)+\ddd\right). \ea We are working
for the choice \ba g(z)=\s{1+z^4/z^4_F}, \label{gfc} \ea which is
asymptotically $AdS_4$ and leads to the IR geometry (\ref{fermig}).
Since we need to perform the numerical analysis below, we will
simplify set $\f{\pi R^2}{2G_N}=1$.

\begin{figure}[ttt]
   \begin{center}
     \includegraphics[height=5cm]{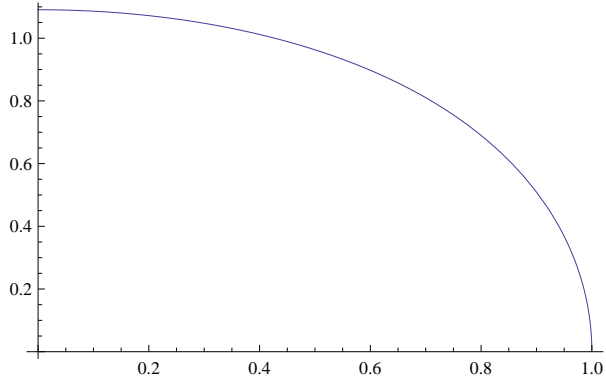}
   \end{center}
   \caption{The shape of the function $r(z)$ for $z_*=1$. }\label{fig:shape}
\end{figure}

\begin{figure}[ttt]
   \begin{center}
     \includegraphics[height=5cm]{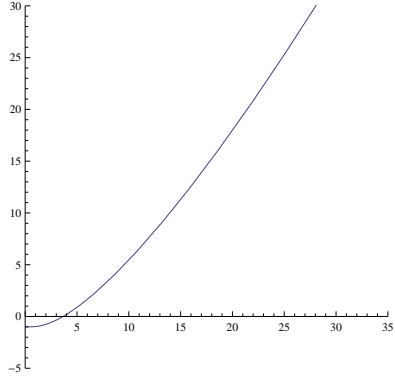}
   \end{center}
   \caption{The finite part of $S_A$ (called $S^{fin}_A$) as a function $l$.}\label{fig:log}
\end{figure}

The explicit form of $r(z)$ is shown in Fig.\ref{fig:shape}.
We plotted the finite part of the entanglement entropy in
Fig.\ref{fig:log} for the choice $z_F=1$. Its behavior is indeed
well approximated by \be S^{fin}_A\simeq 0.50\cdot l\log l-0.60\cdot
l. \label{entl} \ee By looking at the behavior of the coefficients
in (\ref{entl}), we can numerically confirm the dependence of
$z_F\propto k_F^{-1}$ as in (\ref{fsll}).

\section{Null Energy Condition and Specific Heat}

To have a physically sensible gravity theory, we need to require an
appropriate energy condition. In the presence of negative
cosmological constants, we usually impose the null energy condition:
\be T_{\mu\nu}N^\mu N^\nu\geq 0, \label{nec} \ee where $N^\mu$
denotes any null vector. For example, we employ this condition to
prove the holographic c-theorem \cite{cth}.

We find from the Einstein equation that the energy stress tensor $T_{\mu\nu}$ is
simply given by \be T_{\mu\nu}=R_{\mu\nu}-\f{1}{2}g_{\mu\nu}R. \ee
We can choose $N^\mu$ as \be N^t=\f{1}{\s{f(z)}},\ \ \
N^z=\f{\cos\theta}{\s{g(z)}},\ \ \ N^x=\sin\theta, \ee where
$\theta$ is an arbitrary constant.
Then
\ba
T_{\mu\nu}N^{\mu}N^{\nu}
&=& - \frac{
zg(z)f'(z)^2+f(z)(zf'(z)g'(z)+g(z)(4f'(z)-2zf''(z))
}{4zf(z)^2g(z)^2}\sin^2(\theta)
\no
&&
 -\frac{g(z)f'(z)+f(z)g'(z)}{zf(z)g(z)^2}\cos^2(\theta),
\ea
and so the null energy condition \eqref{nec} is satisfied if and only if
\ba &&
g(z)f'(z)+f(z)g'(z)\leq 0, \no &&
zg(z)f'(z)^2+f(z)(zf'(z)g'(z)+g(z)(4f'(z)-2zf''(z)) \leq 0.
\label{condpo} \ea

Let us focus on the IR geometry. If we assume the behavior \be
f(z)\propto z^{-2m}, \ \ \ \ g(z)\propto z^{2n}, \label{irb} \ee
then the conditions  \eqref{condpo} lead
to \be m\geq n.
\label{nuc} \ee

\subsection{Constraints on Specific Heat}

Since the Landau-Fermi liquid has the peculiar property that the
specific heat is linearly proportional to the temperature, it is
helpful to investigate the specific heat. At finite temperature, the
IR geometry (\ref{irb}) is modified into a black hole solution of
the form: \be ds^2=R^2\left(
-z^{-2(m+1)}h(z)dt^2+z^{2(n-1)}\ti{h}(z)^{-1}dz^2+\f{dx^2+dy^2}{z^2}\right),
\label{mecbh}
\ee where we set $R=1$. We define the location of the horizon to be
$z=z_H$. For $\f{z}{z_H}<<1$, the metric should approach
(\ref{irb}) and therefore we have $h(z)\simeq 1$ and $\ti{h}(z)\simeq 1$.
On the other hand, in the near horizon region $z\simeq z_H$, we 
generically have the regular non-extremal horizon and thus they behave like
\be h(z)\sim \ti{h}(z)\sim \f{z_H-z}{z_H}, \ee up to
constant factors. 

Thus we can estimate the temperature as \be T\propto z^{-m-n-1}_H,
\ee by requiring that the Euclidean geometry at $z=z_H$ is smooth.
Since the thermal entropy $S$ is given by the area of horizon we
find \be S \propto \f{V_2}{z^2_H}\propto V_2\cdot T^{\f{2}{m+n+1}},
\ee where $V_2$ is the volume of the dual CFT$_3$.

The specific heat is found as \be C=\f{\de E}{\de T}=T\f{\de S}{\de
T}\propto T^{\f{2}{m+n+1}}. \label{speh} \ee The existence of Fermi
surfaces requires $n=1$ as explained in (\ref{fermig}). In this
case, the null energy condition (\ref{nuc}) leads to\footnote{In
AdS$_5$ gravity duals, we find $C\propto T^{\f{3}{m+n+1}}$ and the
null energy condition requires $m\geq n$. The Fermi surface (or
equally the logarithmic entanglement entropy) corresponds to $n=2$.
Therefore we find the behavior $C\propto T^{\ap}$ with the constraint
$\ap\leq \f{3}{5}$ for the $3+1$ dimensional fermi surfaces.}
 \be
\f{2}{m+n+1}\leq \f{2}{3}. \ee Since the standard Fermi liquid has
the linear specific heat $C\propto T$, classical gravity duals
cannot have the Landau-Fermi liquids. Instead, our specific heat
(\ref{speh}) suggests that they correspond to non-Fermi
liquids\footnote{It might be intriguing to note that a $2+1$
dimensional Kondo breakdown quantum critical point (a non Fermi
liquid) has $C\sim T^{2/3}$ behavior \cite{KB}. Also the same
behavior also appear in a class of non-Fermi liquids in the large N
limit \cite{MMLS}.}.

If the Fermi surface is govern by a two dimensional theory with the
dispersion relation $E\sim k^{\hat{z}}$, then we expect \be C\propto
S\propto T^{\f{1}{\hat{z}}}, \ee which agrees with the analysis in
\cite{Senthil}. The null energy condition requires \be \hat{z}\geq
\f{3}{2}. \label{boundz} \ee

In summary, we found that the Landau-Fermi liquids cannot
holographically be realized in any purely classical
gravity. However, non-Fermi liquids are
not excluded as long as the bound (\ref{boundz}) is satisfied (in
$2+1$ dimension). For such a classical gravity dual,
we expect the presence of order $N^2$
Fermi surfaces, taking the Fermi energy to be order one.

It might be useful to note that the electron stars
\cite{ESa,ESb,ESc} are not included in our definition of Fermi
surfaces with classical gravity duals. Actually, in an electron
star, since there is a smaller number ($<<N^2$) of Fermi surfaces
\cite{ESb}, the entanglement entropy and specific heat contributed
by them only appear as quantum corrections to the classical gravity
calculations (see \cite{ESc} for the analysis of the specific heat).
On the other hand, the free energy is computable in classical
gravity because the chemical potential is taken to be very large as
$O(\s{N})$ in the electron star. The same is also true for the IR
gapped models \cite{SaH}.

\section{Effective Gravity Model}

Finally, we would like to give an effective gravity model which has
the logarithmic behavior (\ref{fermiee}) of the holographic
entanglement entropy, peculiar to systems with Fermi surfaces. We
start with the effective action of Einstein-Maxwell-scalar theory:
\be
S=\f{1}{2\kappa^2}\int\s{-g}\left[(R-2\Lambda)-W(\phi)F^{\mu\nu}F_{\mu\nu}
-\f{1}{2}\de_\mu\phi\de^\mu\phi-V(\phi)\right], \ee where $W(\phi)$
and $V(\phi)$ can be any function. $\Lambda=-\f{3}{R^2}$ is the
negative cosmological constant. We would like to mention that this
system has been extensively studied in \cite{Kiritsis,GuRo,IKNT} and
some of our solutions, especially our scaling solutions discussed in
section 5.1, correspond to their special examples.

\subsection{Equation of Motions and General Solutions}

The equations of motion are summarized as follows. The Einstein
equation reads \ba && R_{\mu\nu}-\f{1}{2}g_{\mu\nu}R+\Lambda
g_{\mu\nu}=\no &&
-\f{W(\phi)}{2}F^{\rho\sigma}F_{\rho\sigma}g_{\mu\nu}+2W(\phi)F_{\mu}^\rho
F_{\nu\rho}-\f{1}{4} \de_\rho\phi\de^\rho\phi
g_{\mu\nu}+\f{1}{2}\de_\mu\phi\de_\nu\phi-\f{1}{2}V(\phi)g_{\mu\nu}.
\label{eins} \ea The Maxwell equation is given by \be
\de_\mu\left(\s{-g}W(\phi)F^{\mu\nu}\right)=0.\label{max} \ee
Finally, the scalar equation of motion reads \be
\f{1}{\s{-g}}\de_\mu(\s{-g}\de^\mu\phi)-\f{\de
V(\phi)}{\de\phi}-\f{\de W(\phi)}{\de \phi}F^{\mu\nu}F_{\mu\nu}=0.
\label{dil} \ee

Our ansatz for solutions is given as follows: The metric is again
given by (\ref{metric}). The dilaton and the gauge potential are
functions of $z$ \be \phi=\phi(z), \ \ \  A_t=a(z). \ee

The Maxwell equation (\ref{max}) leads to
\be
a'(z)=\f{A}{W(\phi)}\s{f(z)g(z)},
\ee
where $A$ is an integration constant.
Notice that $A$ is interpreted as the charge density in the dual quantum field theory.

The Einstein equations (\ref{eins}) can be written as
\ba
&& G_{tt}=\f{R^2}{z^2}f(z)\left(\Lambda+\f{V(\phi)}{2}\right)+\f{A^2z^2}{R^2W(\phi)}f(z)
+\f{f(z)}{4g(z)}\phi'(z)^2,\no
&& G_{zz}=-\f{R^2}{z^2}g(z)\left(\Lambda+\f{V(\phi)}{2}\right)-\f{A^2z^2}{R^2W(\phi)}g(z)
+\f{1}{4}\phi'(z)^2,\no
&& G_{xx}=-\f{R^2}{z^2}\left(\Lambda+\f{V(\phi)}{2}\right)+\f{A^2z^2}{R^2W(\phi)}
-\f{1}{4g(z)}\phi'(z)^2. \label{einem}
\ea

The scalar field equation of motion is given by
\be
\f{z^4}{R^4\s{f(z)g(z)}}\de_z\left(\f{R^2}{z^2}\s{\f{f(z)}{g(z)}}\de_z\phi(z)\right)
-\f{\de V(\phi)}{\de\phi}+\f{2A^2z^4}{R^2W(\phi)^2}\f{\de W(\phi)}{\de\phi}=0. \label{scem}
\ee

Now, for our background $G_{\mu\nu}$ are found to be
\ba
&& G_{tt}=-\f{f(z)(3g(z)+zg'(z))}{z^2g(z)^2},\no
&& G_{zz}=\f{3f(z)-zf'(z)}{z^2f(z)},\no
&& G_{xx}=-\f{z^2g(z)f'(z)^2-4f(z)^2(3g(z)+zg'(z))
+zf(z)(zf'(z)g'(z)+g(z)(4f'(z)-2zf''(z)))}{4z^2f(z)^2g(z)^2},\no \label{gtens}
\ea

From (\ref{gtens}) and (\ref{einem}), the three Einstein equations
are equivalently expressed as follows \ba && V(\phi) \no &&
=\f{z^2g(z)f'(z)^2\!+\!4f(z)^2\!(-6g(z)\!+\!6g(z)^2\!-\!zg'(z))\!+\!zf(z)
\!(zf'(z)g'(z)\!+\!g(z)\!(8f'(z)\!-2zf''(z)\!))
}{4R^2f(z)^2g(z)^2},\no && \f{1}{W(\phi)}=-\f{R^2}{8A^2}\cdot
\f{zg(z)f'(z)^2+f(z)(zf'(z)g'(z)+g(z)(4f'(z)-2zf''(z))}{z^3f(z)^2g(z)^2},\no
&& \phi'(z)^2=-\f{2(g(z)f'(z)+f(z)g'(z))}{zf(z)g(z)}. \label{eomsum}
\ea

It is immediate to find the physical conditions by requiring
$W(\phi)\geq 0$ and $\phi'(z)^2\geq 0$,
and we find that they are exactly equivalent to the null energy conditions
\eqref{condpo}

We can show that (\ref{scem}) is automatically satisfied if we
substitute (\ref{eomsum}) into it. Thus, in principle, we can always
have a full solution to all equations of motion by choosing
$V(\phi)$ and $W(\phi)$ appropriately. Only non-trivial constraint
comes from the no-ghost (and null energy) condition (\ref{condpo}).

\subsection{IR Solution with Logarithmic Entanglement Entropy}

To reproduce (\ref{fermiee}) we need to have \be
g(z)=\f{z^2}{z_F^2}, \label{gz} \ee in the IR region $z>>z_F$. We
also assume the following form of $f(z)$: \be f(z)=k z^{-p}, \ee
where $k$ is a positive constant. In this case, the solution exists
for $p>2$ and it is given by \ba
&& \phi(z)=\s{2(p-2)}\log z, \\
&& V(\phi)=\f{6}{R^2}-\f{(32+12p+p^2)z_F^2}{4R^2z^2},\label{potv}\\
&& W(\phi)=\f{8A^2z^6}{z_F^2p(8+p)R^2}. \label{potw} \ea Therefore
we find \ba && V(\phi)=
\f{6}{R^2}-\f{(32+12p+p^2)z_F^2}{4R^2}e^{-\s{\f{2}{p-2}}\phi},\no &&
W(\phi)= \f{8A^2}{z_F^2p(p+8)R^2}e^{\f{6}{\s{2(p-2)}}\phi}. \ea

It is clear that this solution gets singular in the IR limit
$z=\infty$. In order to show that the singularity at zero
temperature is not problematic, we need to find regular black hole
solutions at finite temperature. We assume the potentials $V(\phi)$
and $W(\phi)$ are given by (\ref{potv}) and (\ref{potw}). Then we
can find the following simple black hole solutions \ba &&
g(z)=\f{z^2}{z_F^2h(z)},   \no && f(z)=k\cdot z^{-p}\cdot h(z),\no
&& \phi(z)=\s{2(p-2)}\log z, \ea where $h(z)$ is given by \be
h(z)=1-M z^{\f{p+8}{2}}. \ee The positive parameter $M$ represents
the mass of the black hole and the horizon is situated at $h(r)=0$.
In this example, we can show that the specific heat $C$ or equally
the thermal entropy $S$ behaves as follows \be C\propto S\propto
\left(\f{T}{z_F\s{k}}\right)^{\f{4}{p+4}}. \label{sph} \ee This
temperature dependence is consistent with the general argument in
(\ref{speh}).

\subsection{Embedding into Asymptotically AdS Solutions}

Finally, we would like to embed the above IR solution into an
asymptotically AdS background. To make the presentation simple, we
concentrate on the $p=3$ case below. Then we can choose \be
g(z)=\s{1+\f{z^4}{z_F^4}},\ \ \ \ f(z)=\f{k\cdot z^{-3}}{1+k\cdot
z^{-3}}. \ee The scalar field is found to be \be
\phi(z)=\s{2}\int^z_0dz\s{\f{z(3z_F^4-2kz+z^4)}{(k+z^3)(z_F^4+z^4)}}.
\ee In order to keep the inside of the square root positive for all
$z$, we need to require \be k<2z_F^3. \label{kf} \ee

The scalar field behaves like
\ba
&& \phi(z)\simeq  \f{2\s{2}}{\s{3k}}z^{3/2}\ \ (z\to 0),\no
&& \phi(z)\simeq \s{2}\log z +\phi_0,
\ea
where $\phi_0$ is a certain constant defined by the above asymptotic behavior.

\begin{figure}[ttt]
   \begin{center}
     \includegraphics[height=5cm]{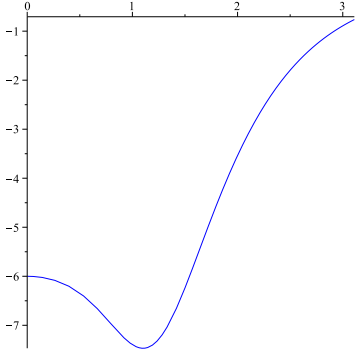}
   \caption{The profile of scalar potential $V(\phi)+2\Lambda$
   for $z_F=k=1$ and $p=3$. 
   The unit of the vertical axis is $1/R^2$. 
   The initial condition of the scalar field is set to be $\phi(0)=0$. }\label{fig:potential}
   \end{center}
\end{figure}

The resulting scalar potential $V(\phi)$ is plotted in Fig.\ref{fig:potential}.
It behaves like
\ba
&& V(\phi)\simeq -\f{9}{8R^2}\phi^2,\ \ \ (\phi\to 0) \label{mass}\\
&& V(\phi)\simeq
\f{6}{R^2}-\f{(32+12p+p^2)z_F^2}{4R^2}e^{-\s{2}(\phi-\phi_0)}\ \ \
(\phi\to \infty). \ea Notice that the tachyonic mass (\ref{mass})
marginally satisfies the BF bound.

\begin{figure}[ttt]
   \begin{center}
     \includegraphics[height=5cm]{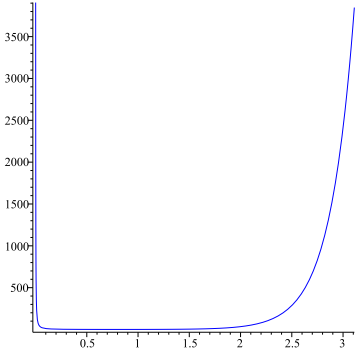}
   \caption{The profile of $W(\phi)$ for $z_F=k=1$ and $p=3$. 
   The unit of the vertical axis is $A^2/R^2$.
   The initial condition of the scalar field is set to be $\phi(0)=0$. }\label{fig:w}
   \end{center}
\end{figure}

The behavior of $W$ is plotted in Fig.\ref{fig:w}. 
It looks like
\ba && W(\phi)\simeq 2^{5}3^{-11/3}\f{A^2k^{4/3}}{R^2}\phi^{-4/3},\
\ \ (\phi\to 0)\no && W(\phi)\simeq
\f{8A^2}{33z_F^2R^2}e^{3\s{2}(\phi-\phi_0)}\ \ \ (\phi\to \infty).
\ea As a function $z$, $W$ is given by \be
W(z)=\f{8A^2(k+z^3)^2(z^4+z_F^4)\s{1+z^4/z_F^4}}{3R^2z^2(9z_F^4+2kz+11z^4)}.
\ee It will be an very intriguing future problem to find a string
theory embedding which has these properties of $V$ and $W$.

In summary, in order to have expected solutions, the potentials
should behave as follows in the limit $\phi\to\infty$ \ba &&
V(\phi)\simeq \f{6}{R^2}-\f{c_1}{R^2}e^{-\s{2}\phi},\no &&
W(\phi)\simeq c_2R^2e^{3\s{2}\phi}, \ea where the dimensionless
constants $c_1$ and $c_2$ are fixed once we decide a microscopic
theory. If we set $k\sim O(z_F^3)$ motivated by (\ref{kf}), then we
find $\phi_0\sim -\s{2}\log z_F$. Thus we find the relation (up to a
numerical constant) \be z_F\sim \f{R}{\s{A}}. \label{relar} \ee

The gauge potential at the AdS boundary $z=0$ is found to be (again
up to numerical constants) \be A_t(0)=\int^{\infty}_0
dz\f{A}{W(z)}\s{f(z)g(z)}\sim \f{R^2}{Az_F^3}\sim \f{1}{z_F}. \ee
This relates the Fermi energy $\mu_F\equiv A_t(0)$ to the charge
density $A$ by $\mu_F\propto \s{A}$. In this case, the specific heat
(\ref{sph}) behaves like \be C\propto
T^{\f{4}{7}}(\mu_F)^{\f{10}{7}}. \ee

\vskip6mm \noindent {\bf Acknowledgments} We are grateful to J.
Bhattacharya, T. Grover, S. Hartnoll, G. Horowitz, N. Iqbal, S-S.
Lee, J. McGreevy, R. Myers, S. Ryu for useful discussions and especially to
H. Liu and S. Sachdev for careful reading of the draft of this paper
and for valuable comments. T.T. thank very much KITP and the organizers of
the KITP program ``Holographic Duality and Condensed Matter Physics'', where
this work has been completed. N.O. is supported by the postdoctoral fellowship
program of the Japan Society for the Promotion of Science (JSPS) and
partly by JSPS Grant-in-Aid for JSPS Fellows No. 22-4554. The work
of T.T. is also supported in part by JSPS Grant-in-Aid for
Scientific Research No.20740132, and by JSPS Grant-in-Aid for
Creative Scientific Research No.\,19GS0219. We are supported by
World Premier International Research Center Initiative (WPI
Initiative), MEXT, Japan.

\newpage
\appendix

\newcommand{\J}[4]{{\sl #1} {\bf #2} (#3) #4}
\newcommand{\andJ}[3]{{\bf #1} (#2) #3}
\newcommand{\AP}{Ann.\ Phys.\ (N.Y.)}
\newcommand{\MPL}{Mod.\ Phys.\ Lett.}
\newcommand{\NP}{Nucl.\ Phys.}
\newcommand{\PL}{Phys.\ Lett.}
\newcommand{\PR}{ Phys.\ Rev.}
\newcommand{\PRL}{Phys.\ Rev.\ Lett.}
\newcommand{\PTP}{Prog.\ Theor.\ Phys.}
\newcommand{\hep}[1]{{\tt hep-th/{#1}}}


\begin{thebibliography}{99}

\baselineskip=10pt


\bibitem{Maldacena}
  J.~M.~Maldacena,
  ``The large N limit of superconformal field theories and supergravity,''
  Adv.\ Theor.\ Math.\ Phys.\  {\bf 2} (1998) 231
  [Int.\ J.\ Theor.\ Phys.\  {\bf 38} (1999) 1113]
  [arXiv:hep-th/9711200];



\bibitem{MIT}
  H.~Liu, J.~McGreevy and D.~Vegh,
  ``Non-Fermi liquids from holography,''
  Phys.\ Rev.\  D {\bf 83} (2011) 065029
  [arXiv:0903.2477 [hep-th]];
     T.~Faulkner, H.~Liu, J.~McGreevy and D.~Vegh,
  ``Emergent quantum criticality, Fermi surfaces, and AdS(2),''
  Phys.\ Rev.\  D {\bf 83} (2011) 125002
  [arXiv:0907.2694 [hep-th]].


\bibitem{Za}
  M.~Cubrovic, J.~Zaanen and K.~Schalm,
  ``String Theory, Quantum Phase Transitions and the Emergent Fermi-Liquid,''
  Science {\bf 325} (2009) 439
  [arXiv:0904.1993 [hep-th]].

\bibitem{Rey}
  S.~-J.~Rey,
  ``String theory on thin semiconductors: Holographic realization
  of Fermi points and surfaces,''
  PTPSA,177,128-142.\ 2009 {\bf 177 } (2009)  128-142.
  [arXiv:0911.5295 [hep-th]].


\bibitem{MITR}
T.~Faulkner, N.~Iqbal, H.~Liu, J.~McGreevy and D.~Vegh,
  ``From black holes to strange metals,''
  arXiv:1003.1728 [hep-th];
 N.~Iqbal, H.~Liu and M.~Mezei,
  ``Lectures on holographic non-Fermi liquids and quantum phase transitions,''
  arXiv:1110.3814 [hep-th].





\bibitem{ESa}
 S.~A.~Hartnoll, J.~Polchinski, E.~Silverstein and D.~Tong,
  ``Towards strange metallic holography,''
  JHEP {\bf 1004} (2010) 120
  [arXiv:0912.1061 [hep-th]];
  S.~A.~Hartnoll and A.~Tavanfar,
  ``Electron stars for holographic metallic criticality,''
  Phys.\ Rev.\  D {\bf 83} (2011) 046003
  [arXiv:1008.2828 [hep-th]];
 S.~A.~Hartnoll,
  ``Horizons, holography and condensed matter,''
  arXiv:1106.4324 [hep-th].



\bibitem{ESb}
  S.~A.~Hartnoll, D.~M.~Hofman and D.~Vegh,
  ``Stellar spectroscopy: Fermions and holographic Lifshitz criticality,''
  JHEP {\bf 1108} (2011) 096
  [arXiv:1105.3197 [hep-th]].


\bibitem{ESc}
  S.~A.~Hartnoll and P.~Petrov,
  ``Electron star birth: A continuous phase transition at nonzero density,''
  Phys.\ Rev.\ Lett.\  {\bf 106} (2011) 121601
  [arXiv:1011.6469 [hep-th]].



\bibitem{SaF}
S.~Sachdev,
``Strange metals and the AdS/CFT correspondence,''
J.\ Stat.\ Mech.\ {\bf 1011}(2010) 11022;
L.~Huijse and S.~Sachdev,
  ``Fermi surfaces and gauge-gravity duality,''
  Phys.\ Rev.\  D {\bf 84} (2011) 026001
  [arXiv:1104.5022 [hep-th]]


\bibitem{ILM}
  N.~Iqbal, H.~Liu and M.~Mezei,
  ``Semi-local quantum liquids,''
  arXiv:1105.4621 [hep-th].

\bibitem{GSW}
J.~P.~Gauntlett, J.~Sonner, D.~Waldram,
``Universal fermionic spectral functions from string theory,''
[arXiv:1106.4694 [hep-th]];
``Spectral function of the supersymmetry current (II),''
[arXiv:1108.1205 [hep-th]].



\bibitem{SaH}
S.~Sachdev,
  ``A model of a Fermi liquid using gauge-gravity duality,''
  Phys.\ Rev.\  D {\bf 84} (2011) 066009
  [arXiv:1107.5321 [hep-th]];
S.~Sachdev,
  ``What can gauge-gravity duality teach us about condensed matter physics?,''
  arXiv:1108.1197 [cond-mat.str-el].



\bibitem{KLM}
  S.~Kachru, X.~Liu and M.~Mulligan,
  ``Gravity Duals of Lifshitz-like Fixed Points,''
  Phys.\ Rev.\  D {\bf 78} (2008) 106005
  [arXiv:0808.1725 [hep-th]].



\bibitem{BoSr}
L.~Bombelli, R.~K.~Koul, J.~H.~Lee and R.~D.~Sorkin,
  ``A Quantum Source Of Entropy For Black Holes,''
  Phys.\ Rev.\ D {\bf 34}, 373 (1986);
  M.~Srednicki,
  ``Entropy and area,''
  Phys.\ Rev.\ Lett.\  {\bf 71}, 666 (1993)
  [arXiv:hep-th/9303048].


\bibitem{EIR}
  J.~Eisert, M.~Cramer and M.~B.~Plenio,
  ``Area laws for the entanglement entropy - a review,''
  Rev.\ Mod.\ Phys.\  {\bf 82}, 277 (2010)
  [arXiv:0808.3773 [quant-ph]].


\bibitem{CaCaR}
 C.~Holzhey, F.~Larsen and F.~Wilczek,
  ``Geometric and renormalized entropy in conformal field theory,''
  Nucl.\ Phys.\  B {\bf 424} (1994) 443
  [arXiv:hep-th/9403108];

 P.~Calabrese and J.~L.~Cardy,
  J.\ Stat.\ Mech.\  {\bf 0406} (2004) P06002
  [arXiv:hep-th/0405152];

P.~Calabrese and J.~Cardy,
  ``Entanglement entropy and conformal field theory,''
  J.\ Phys.\ A  {\bf 42} (2009) 504005
  [arXiv:0905.4013 [cond-mat.stat-mech]].


\bibitem{CaHuR}
  H.~Casini and M.~Huerta,
  ``Entanglement entropy in free quantum field theory,''
  J.\ Phys.\ A  {\bf 42} (2009) 504007
  [arXiv:0905.2562 [hep-th]].



\bibitem{EEF}
M.~M.~Wolf, ``Violation of the entropic area law for fermions'',
Phys.\ Rev.\ Lett.\  {\bf 96} (2006) 010404 [quant-ph/0503219];

D.~Gioev, I.~Klich, ``Entanglement entropy of fermions in any
dimension and the Widom conjecture,'' Phys.\ Rev.\ Lett.\  {\bf 96}
(2006) 100503 [quant-ph/0504151].


\bibitem{SW}
B.~Swingle,
 ``Entanglement Entropy and the Fermi Surface,''
  arXiv:0908.1724 [cond-mat.str-el];
``Conformal Field Theory on the Fermi Surface,''
  arXiv:1002.4635 [cond-mat.str-el].


\bibitem{spin}
Y.~Zhang, T.~Grover, A.~Vishwanath, ``Entanglement entropy of
critical spin liquids'', Phys.\ Rev.\ Lett.\  {\bf 107} (2011)
067202 [arXiv:1102.0350[cond-mat]]. 1


\bibitem{RT}
  S.~Ryu and T.~Takayanagi,
  ``Holographic derivation of entanglement entropy from AdS/CFT,''
  Phys.\ Rev.\ Lett.\  {\bf 96} (2006) 181602
  [arXiv:hep-th/0603001].

\bibitem{RTL}
 S.~Ryu and T.~Takayanagi,
 ``Aspects of holographic entanglement entropy,''
  JHEP {\bf 0608} (2006) 045
  [arXiv:hep-th/0605073].

\bibitem{RTR}
 T.~Nishioka, S.~Ryu and T.~Takayanagi,
  ``Holographic Entanglement Entropy: An Overview,''
  J.\ Phys.\ A  {\bf 42} (2009) 504008
  [arXiv:0905.0932 [hep-th]].

\bibitem{DSY}
W.~Ding, A.~Seidel, K.~Yang, `` Entanglement Entropy of Fermi Liquids via Multi-dimensional Bosonization,'' arXiv:1110.3004.

\bibitem{ANT}
T.~Azeyanagi, T.~Nishioka, T.~Takayanagi,
  ``Near Extremal Black Hole Entropy as Entanglement Entropy via AdS(2)/CFT(1),''
  Phys.\ Rev.\  {\bf D77 } (2008)  064005.
  [arXiv:0710.2956 [hep-th]].

\bibitem{DhKr}
E.~D'Hoker and P.~Kraus,
  ``Magnetic Brane Solutions in AdS,''
  JHEP {\bf 0910} (2009) 088
  [arXiv:0908.3875 [hep-th]];
  ``Charged Magnetic Brane Solutions in AdS5 and the fate of the third law of
  thermodynamics,''
  JHEP {\bf 1003} (2010) 095
  [arXiv:0911.4518 [hep-th]];
 ``Holographic Metamagnetism, Quantum Criticality, and Crossover Behavior,''
  JHEP {\bf 1005} (2010) 083
  [arXiv:1003.1302 [hep-th]].

\bibitem{SWH}
B.~Swingle,
``Highly entangled quantum systems in 3+1 dimensions,''
arXiv:1003.2434.

\bibitem{MFS}
 M.~A.~Metlitski, C.~A.~Fuertes and S.~Sachdev,
 ``Entanglement Entropy in the O(N) model,''
 Phys.\ Rev. {\bf B 80} (2009) 115122 [arXiv:0904.4477].




\bibitem{cth}
  D.~Z.~Freedman, S.~S.~Gubser, K.~Pilch and N.~P.~Warner,
  ``Renormalization group flows from holography supersymmetry and a c
  theorem,''
  Adv.\ Theor.\ Math.\ Phys.\  {\bf 3} (1999) 363;
 R.~C.~Myers and A.~Sinha,
  ``Holographic c-theorems in arbitrary dimensions,''
  JHEP {\bf 1101} (2011) 125.


\bibitem{KB}
C.~Pepin, `` Kondo Breakdown as a Selective Mott Transition in the
Anderson Lattice ,'' Phys.\ Rev.\ Lett.\  {\bf 98} (2007) 206401
[cond-mat/0610846].


\bibitem{MMLS}
D.~Mross, J.~McGreevy, H.~Liu, T.~Senthil, ``
 A controlled expansion for certain non-Fermi liquid metals,''
Phys.\ Rev. {\bf B82} (2010) 04512


\bibitem{Senthil}
T. Senthil, ``Critical fermi surfaces and non-fermi liquid metals,''
Phys.\ Rev.\  {\bf B78 } (2008) 035103 [arXiv:0803.4009 [cond-mat]].



\bibitem{Kiritsis}
  C.~Charmousis, B.~Gouteraux, B.~S.~Kim, E.~Kiritsis and R.~Meyer,
  ``Effective Holographic Theories for low-temperature condensed matter
  systems,''
  JHEP {\bf 1011} (2010) 151
  [arXiv:1005.4690 [hep-th]].


\bibitem{GuRo}
S.~S.~Gubser and F.~D.~Rocha,
  ``Peculiar properties of a charged dilatonic black hole in AdS5,''
  Phys.\ Rev.\  D {\bf 81} (2010) 046001
  [arXiv:0911.2898 [hep-th]].




\bibitem{IKNT}
  N.~Iizuka, N.~Kundu, P.~Narayan and S.~P.~Trivedi,
  ``Holographic Fermi and Non-Fermi Liquids with Transitions in Dilaton
  Gravity,''
  arXiv:1105.1162 [hep-th].







\end{thebibliography}
\end{document}